\pdfoutput=1

\documentclass[10pt,twocolumn,letterpaper]{article}

\usepackage{cvpr}
\usepackage{times}
\usepackage{epsfig}
\usepackage{graphicx}
\usepackage{amsmath}
\usepackage{amssymb}
\usepackage{subfigure}
\usepackage{booktabs}
\usepackage{url}
\usepackage{tabularx}

 \usepackage{color}

\cvprfinalcopy 
\def\cvprPaperID{31}

\begin{document}

\title{Manipulation Detection in Satellite Images Using Deep Belief Networks}

\author{J{\'a}nos Horv{\'a}th, Daniel Mas Montserrat, Hanxiang Hao, Edward J. Delp\\ \\
Video and Image Processing Laboratory (VIPER)\\
School of Electrical Engineering\\
Purdue University\\
West Lafayette, Indiana, USA\\
}

\maketitle

\thispagestyle{empty}

\begin{abstract}
Satellite images are more accessible with the increase of commercial satellites being orbited. 
These images are used in a wide range of applications including agricultural management, meteorological prediction, damage assessment from natural disasters and cartography.
Image manipulation tools including both manual editing tools and automated techniques can be easily  used to tamper and modify satellite imagery.
One type of manipulation that we examine in this paper is the splice attack where a region from one image (or the same image) is inserted (``spliced'') into an image. 
In this paper, we present a one-class detection method based on deep belief networks (DBN) for splicing detection and localization without using any prior knowledge of the manipulations. 
We evaluate the performance of our approach and show that it provides good detection and localization accuracies in small forgeries compared to other approaches.
\end{abstract}

\section{Introduction}

Satellite images can be used in many applications including meteorological measurements, such as precipitation prediction \cite{lebedev_2019}, thunderstorm detection \cite{zhang_2016} and wind speed and direction estimation \cite{sahoo_2019}.
The analysis of satellite images can also be an efficient way to assess regional infrastructure levels \cite{suraj_2017,oshri_2018}, classify crops in agricultural applications \cite{russwurm_2019,brandt_2019},  forest characterization \cite{chauve_2009}, scene classification \cite{amirabbas_2017,shimoni_2008} or to estimate soil moisture  \cite{efremova_2018,alexakis_2017}. 

The number of  commercial satellites is increasing exponentially \cite{ucs} with many of these platforms having advanced imaging sensors. 
These satellites have provided a large number of image datasets available to the public \cite{xia_2019,azimi_2019,yi_2010}, such as Planet Labs or the European Space Agency image datasets \cite{gupta_2019,schmitt_2019}.

Satellite images can be easily forged and manipulated. 
Image and video editing tools such as GIMP \cite{gimp} or Photoshop \cite{photoshop_2016} can be used to create realistic forgeries. 
Machine learning techniques \cite{nam_2018} can quickly manipulate images in an automatic way, without manual edition.
The easy access to image manipulation methods, coupled with an increasing amount of digital content, poses a problem for institutions that rely on satellite imagery. 
Some examples where satellite images were manipulated in order to bias public opinion include the Malaysia Airlines Flight incident \cite{kramer_2016}, the nighttime flyovers of India during the Diwali festivals \cite{byrd_2018} and the fake Chinese bridge spliced images \cite{edwards_2019}.

\begin{figure}[t]
    \vspace{0.25in}
	\centering
	\begin{subfigure}
	\centering
		\includegraphics[width=\columnwidth]{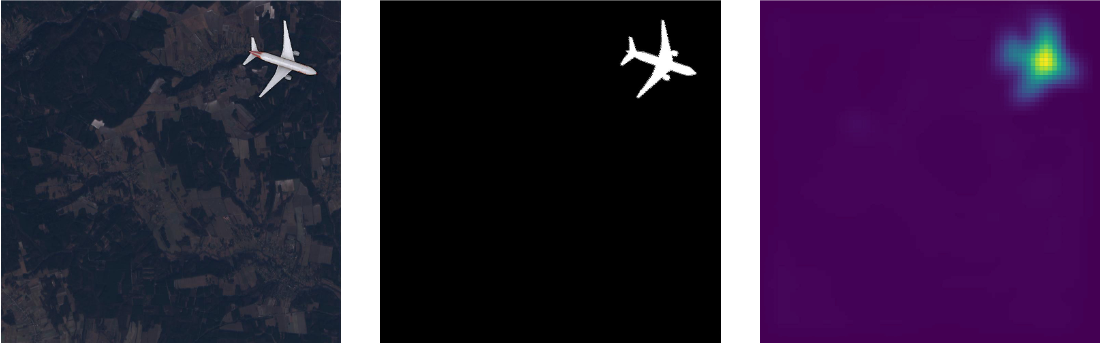}
		\includegraphics[width=\columnwidth]{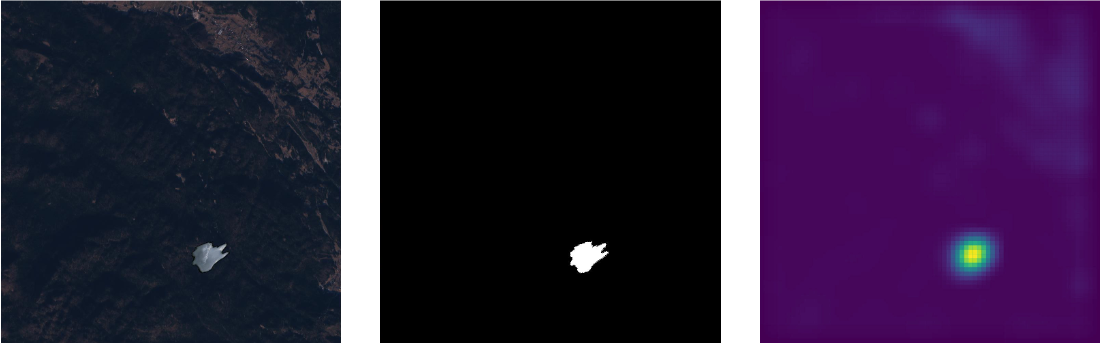}
		\label{fig:large_forg_int}
		\label{fig:medium_forg_int}
	\end{subfigure}
	\caption{Examples of manipulated images (left), manipulation masks (center), and manipulation heatmap (right).}
	\label{fig:method_sum}
\end{figure}

\begin{figure*}[t]
    \centering
    \includegraphics[width=1.8\columnwidth]{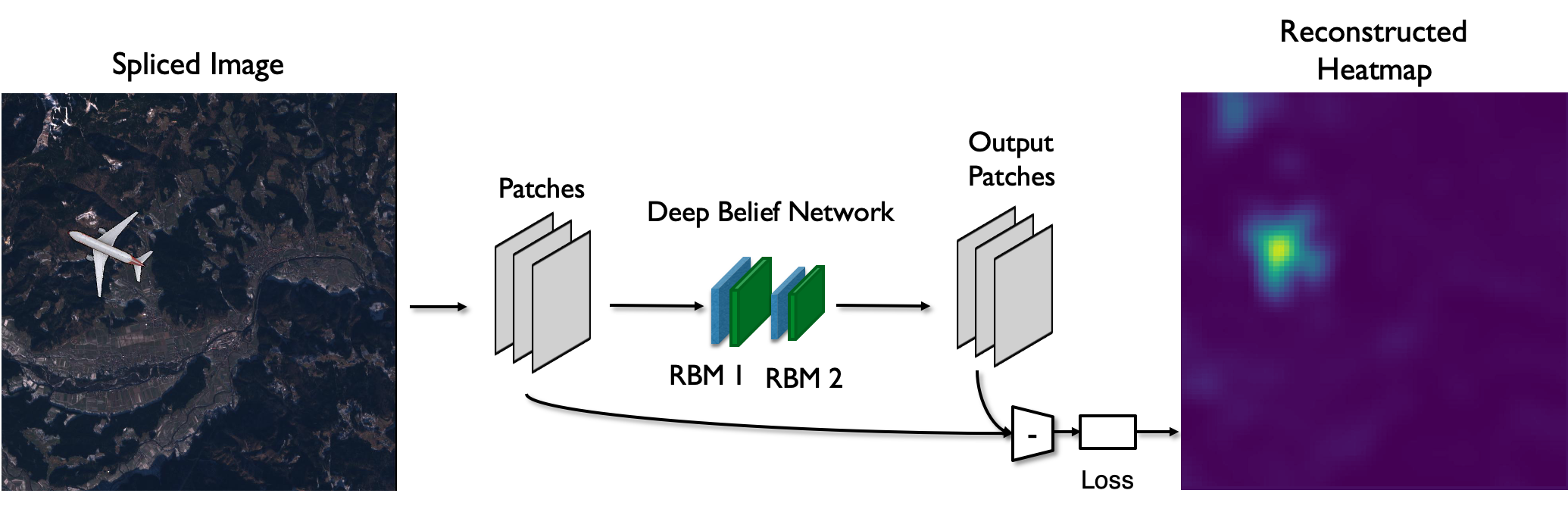} 
    \caption{Method overview for generating the heatmap and the detection score} 
    \label{fig:model_summary}
\end{figure*}

Common manipulations include splicing (cropping and pasting regions from the same or other image sources) and machine learning-based forgeries, typically generated with Generative Adversarial Networks (GANs) \cite{nam_2018}. 
While several methods that validate the authenticity and integrity of images and videos have been proposed \cite{anderson_2011,schetinger_2017, bartolini_2001}, verification of an satellite image authenticity still remains an unsolved problem. 
The wide range of forgery objects and techniques makes their detection challenging. 
Common image forensic methods are often developed for images captured by consumer cameras and fail on satellite imagery as they differ greatly in compression schemes, post-processing, sensors, and color channels. 
There is a need for manipulation detection methods that are accurate regardless of the tampering objects and technologies used for image capturing.

In this paper, we present a method based on Deep Belief Networks (DBN) \cite{hinton_2006} to detect and localize splicing forgeries in satellite images.
Figure \ref{fig:method_sum} shows the results of the proposed method.
The left column shows the manipulated images, the middle column contains the ground truth images of the manipulation and the right column shows the detection results with the accurate locations and shapes of the spliced objects.
A deep belief network is a probabilistic generative model that is composed of multiple layers of Restricted Boltzmann Machines (RBM) \cite{freund_1994}. 
The network learns the distribution of the training data in an unsupervised manner and can be used to detect out-of-distribution data. 
The overview of the proposed method is shown in Figure \ref{fig:model_summary}.
We first generate a set of image patches cropped from the original satellite image.
The deep belief network takes the patches as input and outputs the reconstructed patches.
In this paper, we use two stacked Restricted Boltzmann Machines (RBMs) for the deep belief network. 
At the next step, an error map is computed by combining the difference of the original and reconstructed patches.
The final reconstructed heatmap is obtained by normalizing the error map.
We evaluate our method using a dataset containing splicing forgeries, described in Section \ref{sec:dataset}. 
Additionally, we show that our method can be used as a One-Class classifier and we evaluate it with the MNIST \cite{lecun_2010} dataset.

The main contributions of the paper are that
we present a new technique to generate datasets containing satellite images with splicing manipulations. 
We also introduce a method for splicing manipulation detection and localization using DBNs that does not require any manipulated data during training.
We then evaluate our method with multiple configurations of RBMs.

\section{Related Work} \label{sec:related-work}

The forensics community has described many techniques to detect various types of forgeries and manipulations of images.
Some of these techniques include detecting tampering by finding double-JPEG compression artifacts \cite{barni_2010}, using neural networks with domain adaptation \cite{cozzolino_2018} or using saturation cues \cite{mccloskey_2019}.
Other techniques focus on detecting splicing in images \cite{cozzolino_2015,cozzolino_2020}.
The work presented in \cite{cozzolino_2015}, proposes a new feature-based algorithm to detect splicing in images without any prior information.
They extract expressive features that capture traces left locally by in-camera processing in order to detect manipulations.
The work in \cite{cozzolino_2020} proposes a method to extract the camera model fingerprint in which the scene content is largely suppressed and model-related artifacts are enhanced.
We compare our method performance with both of the previously described splicing detection methods.
Most of these methods fail when used with satellite imagery. 
Because the image acquisition process differs between  ``cameras'' and satellites, methods designed for the former do not transfer properly to the latter. 
These differences include satellite sensor technologies and post-processing steps such as orthorectification, radiometric corrections, and compression.

Several techniques to analyze the integrity of satellite images have been presented using hand-crafted features \cite{ho_2005} and data-driven approaches including both supervised \cite{bartusiak_2019} and unsupervised \cite{sri_2018, horvath_2019} methods. 
Manipulate images are required to use for the supervised techniques during training, but not for unsupervised methods. 
Watermarking techniques have been used for tampering detection \cite{ho_2005}.
The authors in \cite{bartusiak_2019} introduce a supervised approach to detect and localize manipulation in satellite images based on conditional GAN \cite{isola_2017}.
It is trained on original and manipulated images in order to map an input image to a forgery mask. 
The work in \cite{sri_2018} presents a GAN-based method to encode patches extracted from the input image into a low dimensional feature vector. 
Then a one-class support vector machine (SVM) is used to detect if a patch contains forgeries by comparing the distribution of the feature learned from the original image patches.
The Sat-SVDD method \cite{horvath_2019} follows a similar approach by using a modified Support Vector Data Description (SVDD) \cite{tax_2004} to detect splicing forgeries.
Sat-SVDD is a kernel-based one-class classification method that performs minimum volume estimation with a neural network to extract the statistical regularities of the dataset. 
First, the SVDD encodes each image patch to a latent space.
Then, it aims to minimize the distance between each patch and a predefined center point in the latent space. 
By doing so, the model forces the latent vector of original images inside a hypersphere.
At testing, the latent vectors outside the hypersphere are considered as forged data.

In this paper, we use deep belief networks (DBN) \cite{hinton_2006_b}. 
Deep belief networks are composed of stacked layers of restricted Boltzmann machines. \cite{smolensky_1986}.
Boltzmann machines are based on Hopfield networks \cite{hopfield_1982, ackley_1985} which are recurrent neural networks where each input node (or ``visible unit'') is symmetrically interconnected.
These networks can learn the data distribution from a limited set of training examples and reconstruct noisy or incomplete samples through an optimization process of ``energy'' minimization.
Boltzmann machines follow the same structure as Hopfield networks but add a hidden layer connected to the visible layer.
The size and computation of these networks increase exponentially concerning the data dimensionality because all nodes are interconnected.
In order to reduce such complexity, \cite{hinton_2002} introduces a training process based on contrastive divergence minimization.
Furthermore, \cite{hinton_2006} presents the restricted Boltzmann machine (RBM) that reduces the network complexity and training time by removing the connections between nodes (or units) of the same layer.
Many variations of the restricted Boltzmann machine have been presented.
Some examples include the higher-order Boltzmann machine \cite{sejnowski_1986}, the conditional Boltzmann machine \cite{ackley_1985} and the mean field Boltzmann \cite{peterson_1987} machine. 
While regular Boltzmann machines are parametrized with a Bernoulli distribution, many different distributions, including Gaussian have been explored \cite{freund_1994, chu_2019, welling_2005}.
Furthermore, several training methodologies \cite{hinton_2006_b, sutskever_2007, ravanbakhsh_2016} has been presented to reduce the time complexity of the training process. 
More recent work combine DBNs with modern networks such as GANs \cite{huang_2019}.

\section{Dataset} \label{sec:dataset}
\begin{figure}[t]
    \centering
    \includegraphics[width=0.85\columnwidth]{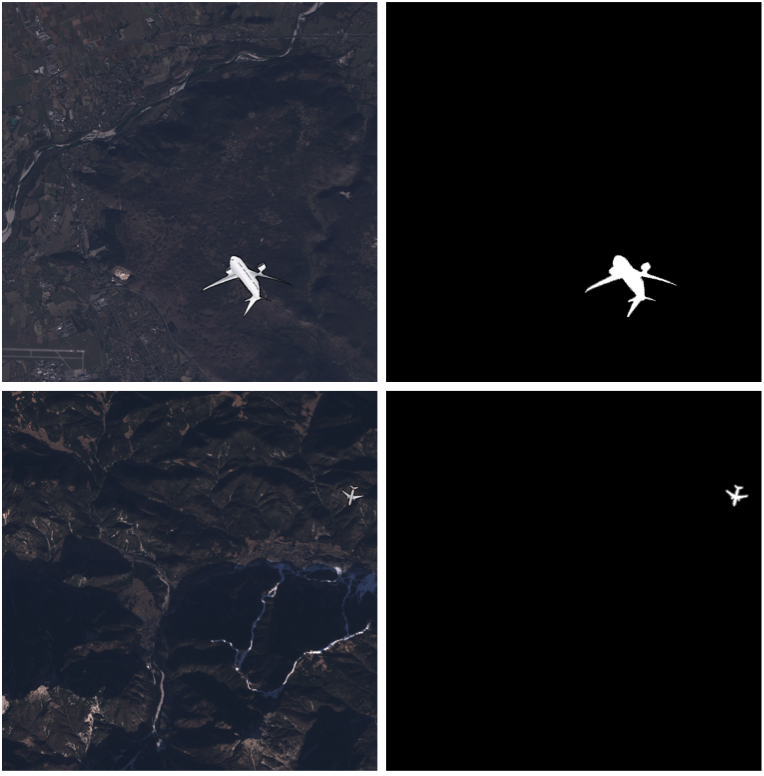}
    \caption{Manipulated dataset examples: two manipulated images with spliced objects (left) and their respective ground truth masks (right).} 
    \label{fig:dataset}
\end{figure}

\begin{figure*}[!htb]
    \centering
    \includegraphics[width=2\columnwidth]{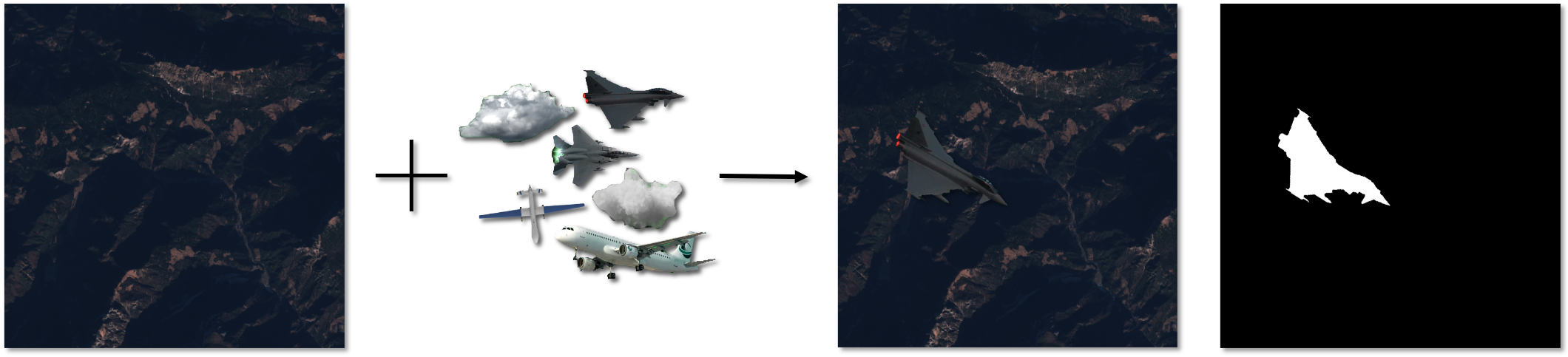}
    \caption{An overview of how the manipulated image dataset is created} 
    \label{fig:dataset-pipeline}
\end{figure*}

In this paper, we use satellite imagery from the region of Slovenia taken from the Sentinel program \cite{Sentinel}.
A total of 293 orthorectified images with an image resolution of $1000\times1000$ pixels are collected.
We use 100 of the 293 orthorectified images to create manipulated images.
19 different objects are spliced into the 100 images generating a total of 500 manipulated images with their corresponding manipulation ground truth masks. 
The 19 objects include clouds, planes, smoke and drones images.
Figure \ref{fig:dataset} shows some examples from the manipulated dataset. 
As shown in Figure \ref{fig:dataset-pipeline}, we splice the objects in different locations, rotation angles and sizes including $16\times16$, $32\times32$, $64\times64$, $128\times128$, and $256\times256$ pixels.
Therefore, the forgery dataset contains a total of 693 images: 193 original images and 500 manipulated images (generated from 100 original images).
We then split the dataset into training and testing sets.
The training dataset consist of 98 original images from the original orthorectified images and the testing dataset consists of 595 images: 95 original images and 500 manipulated images with their corresponding ground truth masks.

\section{Method} \label{sec:method}

\subsection{Splicing Detection}

Our approach has three steps: (1) patch extraction and normalization, (2) patch reconstruction with DBN, and (3) manipulation heatmap and manipulation score estimation as shown in Figure \ref{fig:model_summary}.

Our method starts by cropping overlapping patches of $64\times64$ pixels from the full-resolution input images.
The image patches are obtained by splitting the full-resolution image into a set of overlapping patches with a stride of 32 and 8 pixels during training and testing respectively.
We use a larger stride for training to reduce the computational time and a smaller stride during testing to have a more accurate estimate. 
Finally, each patch is normalized to be in the range of 0 in 1.

Then, the extracted patches are used as input to a DBN.
The DBN learns to reconstruct the input patch. 
We assume that patches from original (non-manipulated) images have different statistical properties than patches containing splicing manipulations.
Therefore, we train the DBN only with original images.
By doing so, the model will properly encode and reconstruct patches from orginal images but will fail with forged patches.

Finally, we compute the mean square error (MSE) between the input patch and the reconstructed patch.
Patches from original (non-manipulated) images will be properly reconstructed, and therefore the reconstruction error will be small.
However, patches with splicing manipulations will be reconstructed poorly leading to a higher MSE.
We create a manipulation heatmap representing the probability of containing a splice manipulation by combining the MSE at each patch.
We average the MSE properly to take into account the overlap within patches.
Once the manipulation heatmap has been estimated, we follow the same approach as in \cite{horvath_2019} to produce a localization mask of an anomaly by thresholding. 

For the anomaly detection task, we compute a detection score to determine the existence of anomaly given a manipulation heatmap.
Following the work presented in \cite{horvath_2019}, we compute a detection score as:

\begin{equation}
\label{eq:threshold}
d(M) = \frac{\max(M)-\mu_M}{\sqrt{\frac{\sum_{x \in M}\left ( x -  \mu_M \right )^{2}}{\max(\left | M \right |)}}},
\end{equation}

Where $M$ is the estimated manipulation heatmap composed by $N$ pixels, $\mu_M = \sum_{x \in M}\frac{x}{N}$ is the average mask value, and $\max(M)$ is the maximum value of $M$.
For a selected threshold $T$, our method considers an input image as manipulated if $d(M)>T$ and original (non-manipulated) otherwise.

\subsection{Network Architecture}

We use a DBN composed by two stacked Restricted Boltzmann Machines (RBM).
The first RBM takes the image patches as input and outputs a feature vector also, referred as hidden representation, of size 2916.
The second RBM takes as input the hidden representation of the first layer and outputs a final hidden representation of the same size as input.
By enforcing a hidden representation with lower dimensionality than the input patch, the network is forced to learn compact features that contain the statistical information of the original image which is used for the image reconstruction task.
Each RBM is specified by the distributions used in their visible (input) and hidden layers. 
We examined three different settings for the RBMs: Gaussian-Bernoulli RBM \cite{chu_2019}, Gaussian-Gaussian RBM \cite{ogawa_2019} and Uniform-Uniform RBM.

The Gaussian-Bernoulli RBM \cite{chu_2019} has been used in previous work \cite{ogawa_2019} and has been shown that it is able to learn features that resemble those learned by the Independent Component Analysis (ICA) algorithm \cite{herault_1984}.
In this RBM the input data is modeled with a Gaussian distribution and the hidden layer is modeled by a Bernoulli distribution.
The conditional probability of the visible layer, given the hidden layer, is defined by the following Normal distribution:

\begin{equation}
\label{eq:normal-1}
p(\mathbf{v}|\mathbf{h})=\mathcal{N}\left (\mathbf{hW}^{T}+\mathbf{a} , \mathbf{I}  \right )
\end{equation}

Where $\mathbf{h}$ are the values of the hidden layer, $\mathbf{W}$ the linear parameters, $\mathbf{a}$ the bias term, and $\mathbf{I}$ the identity matrix.
The conditional probability of the hidden layer given the visible layer is defined by the following Bernoulli distribution:

\begin{equation} 
p(\mathbf{h}|\mathbf{v})=\prod_{i} \sigma (\mathbf{w_i}\mathbf{v}+b_i )^{h_i}(1-\sigma( \mathbf{w_i}\mathbf{v}+b_i ))^{1-h_i}
\end{equation}

Where $w_i$ and $b_i$ are the $i$th row vector of the linear and the bias terms, and $\sigma(.)$ is the Sigmoid function.
By modeling the input with a Normal distribution, it can represent data with a wide range of values (such as images).
The Bernoulli distribution enforces a binary (and therefore sparse) representation of the hidden layer \cite{karakida_2016}.

The Gaussian-Gaussian RBM \cite{ogawa_2019} uses Normal distributions to model both visible and hidden layers.
Such method has been shown to be able to learn features that resemble Principal Component Analysis (PCA) features \cite{pearson_1901}.
The conditional probability of the visible layer, given the hidden layer, is represented in the same way as in Equation \ref{eq:normal-1}. 
The conditional probability of the hidden layer given the visible layer is:

\begin{equation}
p(\mathbf{h}|\mathbf{v})=\mathcal{N}\left (\mathbf{Wv}+\mathbf{b} , \mathbf{I}  \right )
\end{equation}

Where $\mathbf{v}$ are the values of the visible layer, $\mathbf{W}$ the linear parameters, $\mathbf{b}$ the bias term, and $\mathbf{I}$ the identity matrix.
By modelling both visible and hidden layer with a Normal distribution, the network is able to learn a dense representation of the images. 

Additionally, we explore using another setting, the Uniform-Uniform RBM. 
In this network, both visible and hidden layer are modeled with uniform distributions.
The conditional probability of the visible layer, given the hidden layer is:
\begin{equation}
p(\mathbf{v}|\mathbf{h})=U\left ( \lambda_{1} \left (  \mathbf{\mathbf{hW}}^{T}+\mathbf{\mathbf{a}}\right ), \lambda_{2} \left (   \mathbf{hW}^{T}+\mathbf{a}\right )  \right )
\end{equation}

The conditional probability of the hidden layer given the visible layer is:
\begin{equation}
p(\mathbf{h}|\mathbf{v})=U\left ( \lambda_{1} \left (  \mathbf{Wv}+\mathbf{b}\right ),\lambda_{2} \left (   \mathbf{Wv}+\mathbf{b}\right )  \right )
\end{equation} 

Where $\mathbf{v}$ are the values of the visible layer, $\mathbf{h}$ are the values of the hidden layer, $\mathbf{W}$ the linear parameters, and $\mathbf{a}$ and $\mathbf{b}$ the bias terms. We empirically select the values $\lambda_{1}$ and $\lambda_{2}$ that provide a low reconstruction error on the training set. In this work we use $\lambda_{1}=\frac{3}{4}$ and $\lambda_{2}=\frac{5}{4}$.

\subsection{One-Class Classifier}
The proposed method can be used for the one-class classification task.
In a one-class classification problem, the classifier learns important features to identify a known target class based on a training set containing only images of such class.
During testing, the classifier distinguishes between images of the target class and images from other unknown previously unseen classes, based on the features learned during training.
Such an approach is especially useful for outlier and anomaly detection.

In order to evaluate our method as a one-class classifier, we train and test it to perform as handwritten digit recognition with the MNIST dataset \cite{lecun_2010}. In this scenario, the network only has access to images of one target digit during training. Then, the network has to distinguish between the target digit and other digits never seen during training.
When performing a one-class classification, we do not divide the image within patches but use the DBN to reconstruct the complete input image. 
In this setting, the manipulation heatmap $M$ and the manipulation score $d(M)$ are not estimated. 
Instead, the reconstruction error is directly used to classify an image as the known class or not. 
If the reconstruction error is below a chosen threshold $T$, the image is classified as the known class. 
Section \ref{sec:experiment_setup} presents some experiments of one-class classification with MNIST dataset \cite{lecun_2010}.

\section{Experimental Results} \label{sec:experiment_setup}

We evaluate our method as a splicing detection and localization method with the dataset presented in Section \ref{sec:dataset}. 
We also examine the proposed method of the one-class classification task with the MNIST \cite{lecun_2010} dataset that contains images of handwritten digits from 0-9.

\begin{figure*}[!htb]
	\centering
	\begin{subfigure}
	\centering
		\includegraphics[width=0.83\textwidth]{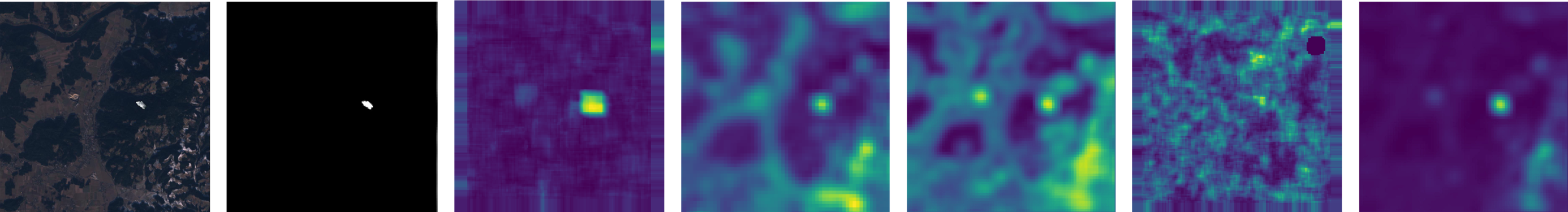}
		\centering
		\includegraphics[width=0.83\textwidth]{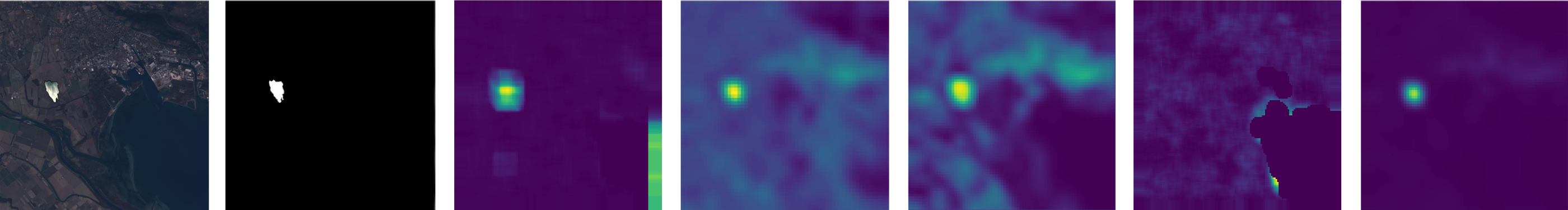}
        \centering
        \vskip 0.07in
		\includegraphics[width=0.83\textwidth]{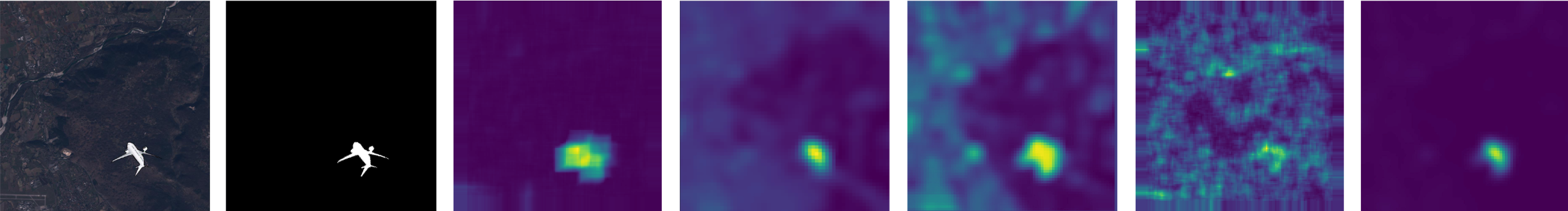}
		
		\label{fig:small_forg}
		\label{fig:medium_forg}
		\label{fig:large_forg}
	\end{subfigure}
	\caption{From rows left to right: input images, ground truth masks, Cozzolino \textit{et al} ~\cite{cozzolino_2015}, Yarlagadda \textit{et al} ~\cite{sri_2018}, Horv{\'a}th \textit{et al} ~\cite{horvath_2019}, Cozzolino \textit{et al} ~\cite{cozzolino_2020} and UU-DBN.}
	\label{fig:results}
	\vskip 0.1in
\end{figure*}

We train and test our DBN based on Uniform-Uniform RBMs (UU-DBN) with the complete training splicing manipulation datasets and compare it with previous work. 
Note that the UU-DBN is trained only with original images and no manipulated images were used during training. 
We compute the ROC curves for detection and Precision/Recall curves for localization tasks by changing the threshold used to the estimated manipulation mask and manipulation score. 
Note that when evaluating the localization performance, there is a larger number of original than manipulated pixels. 
Therefore we use the AUC of the Precision/Recall curve, which is a balanced metric, to evaluate the localization performance.
Table \ref{table:detection_satellite} presents our results compared with previous methods. 
Different ROC and Precision/Recall  are shown for each of the different sizes of the splice objects used. 
In other words, ROC$_{16}$ is the ROC  for manipulated images with spliced objects of size $16\times16$. 
We can observe that UU-DBN provides similar or better results compared to previous methods.
It is able to correctly detect and localize small splicing forgeries.
The method provides a better P/R localization scores, specifically for forgeries with sizes from 16 $\times$ 16 pixels to 128 $\times$ 128 pixels.
Figure \ref{fig:results} provides some visual examples of the estimated manipulation heatmaps for each of the methods. 
The estimated manipulation heatmaps show that the method is able to properly distinguish between splicing manipulations and the background image.

Additionally, we train and test four different DBNs as one-class classifiers with MNIST images. 
We use DBNs based on RBNs with Bernoulli-Bernoulli (BB-DBN), Gaussian-Bernoulli (GB-DBN), Gaussian-Gaussian (GG-DBN) and Uniform-Uniform (UU-DBN) distributions. 
We compare our results with previous one-class classification methods. 
Table \ref{table:detection_mnist} shows the AUC scores for each method. 
We can observe that the presented method provides competitive results with previous methods and that a DBN based on Uniform-Uniform distribution provides higher accuracy. The Gaussian-Bernoulli, Gaussian-Gaussian and Uniform-Uniform DBNs provide a similar AUC score while the Bernoulli-Bernoulli DBN provides a lower score. The binary representation used in the input and hidden layers of the Bernoulli-Bernoulli DBN provides features less flexible that in turn provide worse classification performance.

\begin{table*}[!htb]
	\centering
	\caption {AUC scores (\%) for the detection and localization task (ROC and P/R metrics). The subscript denotes the manipulation size.
	Best performing methods are \textbf{bold}.}
	\label{table:detection_satellite}
	\begin{tabular}{@{}lccccc@{}}
	    & \multicolumn{5}{c}{Detection}  \\
		\toprule
					& Cozzolino \textit{et al} ~\cite{cozzolino_2015} & Yarlagadda \textit{et al} ~\cite{sri_2018} & Horv{\'a}th \textit{et al} ~\cite{horvath_2019} & Cozzolino \textit{et al} ~\cite{cozzolino_2020} & UU-DBN \\
		ROC$_{16}$  & $49.7$                            & $50.7$                      & $45.3$ & $47.7$ &  \textbf{58.2} \\
		ROC$_{32}$  &  $50.4$                          & $59.6$                     & $57.7$   & $47.2$ &   \textbf{68.5}\\
		ROC$_{64}$  & $68.6$                            & $75.9$                       & $80.0$  & $50.8$ &  \textbf{82.6} \\
		ROC$_{128}$ & $84.8$                            & $81.5$                        & $87.4$  & $56.7$ &  \textbf{88.3} \\ 
		ROC$_{256}$ & $86.2$                            & $83.8$                        & \textbf{89.9}  & $55.4$ &  $89.6$ \\ \bottomrule
	\end{tabular}
	\vskip 0.3in
	\begin{tabular}{@{}lccccc@{}}
	    & \multicolumn{5}{c}{Localization} \\
		\toprule
					& Cozzolino \textit{et al} ~\cite{cozzolino_2015} & Yarlagadda \textit{et al} ~\cite{sri_2018} & Horv{\'a}th \textit{et al} ~\cite{horvath_2019} & Cozzolino \textit{et al} ~\cite{cozzolino_2020} & UU-DBN \\
		P/R$_{16}$  & $0.0$     & $0.0$     & $0.1$ & $0.0$ & \textbf{7.5} \\
		P/R$_{32}$  & $0.5$     & $0.3$     & $1.4$ & $0.1$ & \textbf{13.3}\\
		P/R$_{64}$  & $7.8$     & $2.5$     & $18.1$ & $2.5$ & \textbf{31.7}\\
		P/R$_{128}$ & $31.2$     & $18.3$     & $34.4$ & $4.6$ & \textbf{40.5}\\
		P/R$_{256}$ & $48.5$     & $37.8$     & \textbf{55.7} & $7.8$ & $48.8$\\ \bottomrule
	\end{tabular}
	\vskip 0.1in
\end{table*}

\begin{table*}[!htbp]
	\centering
	\caption {AUC scores (\%) for the detection task Receiver operating characteristic (ROC), 
	The best performing DBN configuration is marked in \textcolor{blue}{blue}. The methods with best performance are  \textbf{bold}.}
	\label{table:detection_mnist}
	\begin{tabular}{@{}lcccc|ccccc@{}}
		\toprule
		Class & BB-DBN & GB-DBN  & GG-DBN & UU-DBN & OCSVM  & DCAE & ONE-CLASS  & RCAE \\ & & & & & /SVDD  \cite{scholkopf_2001} & \cite{masci_2011} & DEEP SVDD \cite{lukas_2018} & \cite{zhou_2017}  \\ \midrule
		0  & $97.0\pm0.1$ & $99.1\pm0.0$ & $99.1\pm0.0$ & \textcolor{blue}{$99.3\pm0.1$} & $97.0\pm0.7$ & $97.6\pm0.7$ & $97.2\pm1.2$ & \textbf{99.8}$\pm$\textbf{0.0}\\
		1  & $99.7\pm0.0$ & \textcolor{blue}{\textbf{99.8}$\pm$\textbf{0.0}} & \textcolor{blue}{\textbf{99.8}$\pm$\textbf{0.0}} & \textcolor{blue}{\textbf{99.8}$\pm$\textbf{0.1}} & $98.6\pm0.8$ & $98.3\pm0.6$ & $97.1\pm1.5$ & \textbf{99.8}$\pm$\textbf{0.0}\\
		2  & $79.9\pm0.3$ & $86.4\pm0.2$ & $86.5\pm0.1$ & \textcolor{blue}{$89.5\pm0.1$} & $78.9\pm2.1$ & $85.4\pm2.4$ & $97.3\pm0.8$ & \textbf{98.8}$\pm$\textbf{0.6} \\
		3  & $85.2\pm0.2$ & $91.3\pm0.1$ & $91.4\pm0.1$ & \textcolor{blue}{$93.5\pm0.1$} & $84.3\pm2.6$ & $86.7\pm0.9$ & $97.3\pm1.0$ & \textbf{99.3}$\pm$\textbf{0.0}\\
		4  & $90.0\pm0.2$ & $95.1\pm0.1$ & $95.2\pm0.1$ & \textcolor{blue}{$96.1\pm0.0$} & $93.6\pm1.5$ &  $86.5\pm2.0$ & $97.3\pm1.1$ & \textbf{99.2}$\pm$\textbf{0.0}\\
		5  & $86.2\pm0.2$ & $92.9\pm0.1$ & $93.0\pm0.1$ & \textcolor{blue}{$94.7\pm0.1$} & $74.6\pm4.5$ & $78.2\pm2.7$ & $97.1\pm1.2$ & \textbf{99.2}$\pm$\textbf{0.0}\\
		6  & $94.9\pm0.1$ & $98.3\pm0.0$ & $98.3\pm0.0$ & \textcolor{blue}{$98.7\pm0.1$} & $95.4\pm1.2$ & $94.6\pm0.5$ & $96.7\pm1.3$ & \textbf{99.8}$\pm$\textbf{1.0}\\
		7  & $93.8\pm0.1$ & $95.6\pm0.1$ & $95.8\pm0.1$ & \textcolor{blue}{$96.4\pm0.0$} & $91.9\pm1.5$ & $92.3\pm1.0$ & $97.4\pm1.0$ & \textbf{99.2}$\pm$\textbf{0.1}\\
		8  & $78.4\pm0.1$ & $84.4\pm0.1$ & $85.1\pm0.1$ & \textcolor{blue}{$87.4\pm0.1$} & $87.9\pm1.5$ & $86.5\pm1.6$ & $97.2\pm0.6$ & \textbf{98.5}$\pm$\textbf{2.0}\\
		9  & $90.0\pm0.1$ & $94.5\pm0.1$ & $94.5\pm0.1$ & \textcolor{blue}{$95.3\pm0.1$} & $93.3\pm1.2$ & $90.4\pm1.8$ & $96.6\pm1.5$ & \textbf{99.0}$\pm$\textbf{1.3}\\\hline
	\end{tabular}
	\vskip 0.1in
\end{table*}

\section{Conclusions}\label{sec:conclusion}
Satellite image manipulation detection is a challenging task due to the wide range of manipulations that can be present and the large variety of imaging technology used in orbiting satellites. 
In this paper we present an unsupervised splicing manipulation and one-class classification method based on deep belief networks. 
Without any prior knowledge from the manipulation information during the training process, the method provides competitive results compared to the previous work, especially with small splicing manipulations. 
We also evaluate multiple configurations of our network in a one-class classification framework providing competitive results compared to the common one-class classification methods.
To improve the performance of our proposed method we plan to examine more different structures of Deep Belief Networks and Restricted Boltzmann Machines.
Furthermore, we plan to examine other type of generative methods such as variational autoencoders.

\section{Acknowledgment}
This material is based on research sponsored by DARPA and Air Force Research Laboratory (AFRL) under agreement number FA8750-16-2-0173. 
The U.S. Government is authorized to reproduce and distribute reprints for Governmental purposes notwithstanding any copyright notation thereon. 
The views and conclusions contained herein are those of the authors and should not be interpreted as necessarily representing the official policies or endorsements, either expressed or implied, of DARPA and Air Force Research Laboratory (AFRL) or the U.S. Government.

Address all correspondence to Edward J. Delp, ace@ecn.purdue.edu .
{\small
\bibliographystyle{IEEEbib}
\bibliography{egbib}
}

\end{document}